\documentstyle[12pt,psfig]{article}
\renewcommand{\baselinestretch}{1.4}
\newcommand{\beeq}{\begin{equation}}
\newcommand{\eneq}{\end{equation}}
\newcommand{\beeqar}{\begin{eqnarray}}
\newcommand{\eneqar}{\end{eqnarray}}
\begin{document}

\hfill{KEK Preprint 95-14}

\hfill{KEK-TH-436}

\hfill{KEK-CP-27}
\vskip 0.5in
\begin{center}
{\Large \bf Finite Size Effects for the Ising Model Coupled to
2-D Random Surfaces}

\vspace{4mm}

\renewcommand
\baselinestretch{0.8}\vspace{4mm}
{\sc N. D. Hari Dass\footnote{e-mail:
dass@theory.kek.jp$^{,}$

$^{\rm a}$On leave of
absence from The Institute of Mathematical Sciences, Madras,
India}$^{,\;\rm a}$
, B. E. Hanlon\footnote{e-mail: irish@theory.kek.jp; JSPS research
fellow.}
and T. Yukawa\footnote{e-mail: yukawa@theory.kek.jp}}
\\
{\it
KEK Theory Group, \\
1-1 Oho, \\
Tsukuba, \\
Ibaraki 305, \\
Japan}

\renewcommand
\baselinestretch{1.4}

\vspace{5mm}

\end{center}

\begin{abstract}
Finite size effects for the Ising Model coupled to two dimensional
random surfaces
are studied by exploiting the exact results from the 2-matrix models.
The
fixed area partition function is numerically calculated with arbitrary
precision by developing an efficient algorithm for recursively
solving the
quintic equations so encountered. An analytic method for studying
finite
size effects is developed based on the behaviour of the free energy
near
its singular points. The generic form of finite size corrections so
obtained are seen to be quite different from the phenomenological
parameterisations used in the literature. The method of
singularities is
also applied to study the magnetic susceptibility.
A brief discussion is
presented on the implications of these results to the problem of a
reliable
determination of string susceptibility from numerical simulations.
\end{abstract}
\newpage
\section{Introduction}

Numerical
simulations are important
for the study of nonperturbative effects
which are difficult to handle analytically.
In such simulations one necessarily has to
work with systems of finite size. The effects of
finite size
manifest themselves as systematic errors in measurements. These
corrections are
generically difficult to estimate as they involve dynamical details.
A practical approach is thus often resorted to by applying
phenomenological parametrisations to
finite size effects{\cite A}.

In this paper we address the  question of
finite size corrections by considering the
model of Ising spins coupled to two dimensional random surfaces.
Since the Ising spin case, with central charge
$c_{M}=1/2$, is exactly solvable by the method of matrix models{\cite
{B,C,D}} it
provides an important test case for the efficacy of
 numerical simulations.
 An understanding
of the nature of finite size corrections in this case may be useful
in  probing the unknown
region beyond $c_{M}=1$.

We approach this issue numerically and analytically by
exploiting the parametric
solution of the two matrix models{\cite {C,D}}.
In the numerical approach the inherent quintic equations are
solved recursively from which the
fixed area partition sums  are extracted for various spin couplings.
We also extend this approach to the case of nonzero
magnetic field and determine the magnetic susceptibility scaling laws.
The numerical analysis is augmented by an analytic analysis of the
free energy
about its singular points. From this we suggest a general ansatz for
the form of finite
size corrections.
We
apply our results to different estimates of the string
susceptibility and to
the minbu technique{\cite E},
demonstrating the difference with the phenomenological estimates
previousy employed{\cite A}. The method of singularities is
also applied to
the problem of the finite size behaviour of the magnetic
susceptibility.
It is found that these agree with the usual
finite size scaling laws{\cite F}.

\section{Numerical Solution of the Two Matrix Model}

Matrix models are solved by the method of orthogonal polynomials
whereby a
parametric solution for the free energy is obtained.
The physical couplings
are related
to the
parameter so introduced
by what  we shall call a ``constraint". From
such a solution the relevant critical exponents can be derived.
We will restrict our analysis to surfaces topologically equivalent to
$S^{2}$.

For pure gravity (the one matrix model) the constraint equations
are quadratic and cubic for the case of quartic and cubic interactions
respectively. The constraints are thus solvable by radicals allowing
the constraint to be inverted. From this it is possible to derive a
closed form for  the free energy in the coupling $g$.
For a quartic interaction the series solution for the
free energy is given by{\cite B}:
\beeq
{\cal F} = - \sum_n {(-12g)^n (2n-1)!\over{n!(n+2)!}} \;\; ,
\eneq
which can be written in the form
${\cal F} =  \sum_n {\cal Z}_n g^n$,
${\cal Z}_n$ being the fixed area partition sum for random surfaces.
By exploiting the asymptotic
expansion for the gamma function
it is a simple matter to explicitly derive
 the asymptotic form of the fixed area partition sum:
\beeq
{\cal Z}_n\sim -{(-1)^{n}(48)^{n}\over \sqrt {4\pi}}
n^{-7/2}(1-{25\over 8n}.....) \;\; .
\label{orig}
\eneq
The leading order behaviour corresponds to that
originally predicted by KPZ{\cite H} and has the generic
form
${\cal Z}_n \sim \exp(\mu n) n^{-b}$,
where the string susceptibility is defined as
$\gamma = -b +3$,
and $\mu$ is the cosmological constant.
Similar results can be obtained in the case of cubic interactions as
well as for models with tadpole and/or self energy contributions
removed{\cite B}.

To represent models in which Ising spins have been coupled to the,
discretised,
random surface a two matrix model is required. The method of solution
follows as before. Again the free energy
is given as a parametric solution in
terms of a parameter $z$.
However, we now have the difficulty of a
 constraint that is a quintic{\cite G}, which for quartic
interactions is given by ( where $c=\exp (-2\beta )$ and $\beta=1/T$):
\beeq
g(z) \equiv {z\over(1-3z)^2} -c^2z +3c^2z^3 = g \;\; ,
\label{im}
\eneq
The general quintic is not solvable by radicals. As such a simple
closed form for the free energy cannot be derived, as was possible in
the case of pure gravity, and consequently the asymptotic form
of the fixed area partition sums cannot be easily
extracted.

In
this section we will describe our numerical method of solution to
this problem where
we seek a power series expansion for the free energy in terms of $g$.
We will
describe in detail the case for quartic interactions.
The solution in
the quartic case contains all the essential elements,
while the cubic case{\cite D} is both algorithmically and
algebraically more cumbersome.

To avoid the logarithm in the solution for the free energy we study
${d{\cal F}/ dg}$ which is equivalent in this instance to
${\partial {\cal F}/
{\partial g}}$ yielding:
\beeq
{\partial {\cal F}\over {\partial g}}= [-{1\over {2g}}+
{1\over g^2}\int_0^z{dt\over t}g(t) - {1\over g^2}\int_0^z{dt\over t}
g(t)^2] \;\; .
\label{eqn}
\eneq
We wish to represent this as a simple polynomial in $z$. To remove the
inverse powers of $(1-3z)$ which arise we employ the quintic
constraint (\ref{im}). ${\partial {\cal F}/ {\partial g}}$ is then a
polynomial in $z$ of degree 8 with coefficients depending on the
physical parameters $c$ and $g$. This can further be reduced to a
polynomial of degree 5 by using the constraint to express $z^{8},
z^{7}$ and $z^{6}$ as quintic polynomials in $z$.

To solve the quintic (\ref{im}) numerically we develop an efficient
algorithm
to handle the multiple sums inherent in a power series solution of this
expression.
We require an algorithm which avoids
recalculation. We begin by explicitly expressing each power of $z$ as
a power series in $g$:
\beeq
z = \sum a_1(n)~g^n;~~~~z^2 = \sum~a_2(n)~g^n;.....
{}~~z^5 = \sum a_5(n)~g^n \;\; .
\eneq
Using (\ref{im}), and the fact that $a_{2}(1)=a_{3}(1)=a_{4}(1)=a_{5}
(1)=0$
it is a simple matter to extract $a_{1}(1)=1/ (1-c^2)$. Indeed, a
recursion relation for $a_{1}(n)$ for $n \geq 2$ can be derived from
(\ref{im}) which has the general form:
\beeq
a_1(n) =
\xi_1~a_1(n-1)+\xi_2~a_2(n)+\xi_3~a_2(n-1)+\xi_4~a_3(n)+\xi_5~a_4(n)+
\xi_6~a_5(n) \;\; .
\eneq
where the $\xi$'s are the coefficients appearing in the quintic.
The $a_{2}(n)$'s, $a_{3}(n)$'s etc can ultimately be described in
terms of the $a_{1}(n)$'s. For each power of $z$ the coefficients are
dependent on a calculation of those of lower orders such that for
example
\beeq
a_2(N) = \sum_1^{N-1}a_1(n)\cdot a_1(N-n) \;\; .
\eneq
Thus in order to calculate
$a_2(N)$ we need to find $a_1(k)$ up to
$k=N-1$. Likewise,
$a_3(N)=\sum_1^{N-1}a_1(n)\cdot a_2(N-n)$ implies that a
calculation of $a_3(N)$ requires $a_{2}(N-1)$, which requires
$a_1(N-2)$ etc.

By this approach the calculation time is significantly reduced ( in
this case the calculation time goes as $\sim N^{2}$ as opposed to $\sim
N^{6}$ for a naive approach to multiple sums). Substituting into
(\ref{eqn}) and integrating with respect to $g$ we can recover the
fixed area
partition sums up to large orders in a convenient time.

We must, however, be careful to consider the exponential growth in
the fixed
area partition sums due to the cosmological term.
To
overcome this problem we estimate the cosmological term
by studying the
fixed area partition sums up to areas allowed by machine limits.
The fixed area sums are then scaled by scaling all
the $a(N)$ by this estimate, so that
\beeq
a(N) \rightarrow
 e^{-\mu N}a(N)\;\; ; \;\; \xi_1 \rightarrow
e^{-\mu}\xi_1\;\; , \;\;
\xi_3 \rightarrow e^{-\mu}\xi_3 \;\; .
\eneq
We found that in this way sufficiently  accurate estimates of the
cosmolgical
constant could be obtained for relatively small values of
$n(\simeq 200)$ which could then be used to  extract the
scaled fixed area
partition sums for even large $n\simeq 100000$ easily.
This makes it possible to investigate regions corresponding to
those typical of numerical simulations
via dynamical triangulation, and far beyond.

To investigate the magnetic susceptibility we follow the same general
prescription as above. With the introduction of a magnetic field
two different
coupling constants arise.
In a perturbative solution to lowest order in $H$ the
constraint (\ref{im}) becomes
\beeq
g(z,H) = {z\over (1-3z)^2} -c^2z+ 3c^2z^3 +
{z^2H^2\over {(1-3z)^2(1+3z)^2}}
\;\; .
\label{im2}
\eneq
Since we require solutions in the limit $H\rightarrow ~0$ the
power series
expansions for
$z,z^{2},...$ etc in $g$ remain defined as before by (\ref{im}).
The high and low
temperature phases corresponding to the singularities of the
free energy with
respect to $g$ are defined by the expression $g^{\prime}(z,H=0)=0$.
This has five
solutions of which only two are physical for  $0 <
c < 1$:
\beeq
z_0 = -1/3~~~({\rm Low \;\; temp.\;\; phase})\;\;\; {\rm and} \;\;
\; z_0 = z_0(c)~~~
({\rm High\;\; temp.\;\; phase})
\label{zo}
\eneq
The critical temperature
is at
$c=1/4$.

The magnetic
susceptibility is essentially given by the second
derivative of $\cal F$ with respect to $H$ at $H=0$. In
particular we find
\beeq
K(z)\equiv {\partial^2 {\cal F}\over{\partial  H^2}}|_{H=0}
 = \int_0^{z}
{dt\over t}[{g(t,H)\over g^2}-{1\over g}]
{\partial^2 g\over{\partial H^2}}|_{H=0} \;\; .
\label{mag}
\eneq
We can express this as before as a power series in $g$.
We thus have an expansion of the form
$K(z) = \sum K_n~g^n $.
The fixed area
magnetic susceptibility is then given by
\beeq
\chi_n = {K_n\over nZ_n} \;\; .
\eneq
This is valid at or above the critical temperature. Below the critical
temperature we must account for the spontaneous ordering of spins
so that the
fixed area magnetic susceptibilty is given by
\beeq
\chi_n = {K_n\over nZ_n} -n<\sigma>^2 \;\; ,
\eneq
where $<\sigma >$ is the spontaneous magnetization.

\section{Results from the Numerical Analysis}

We first exhibit the finite size effects
in string susceptibility. For the case at hand $\gamma$ is
known exactly:
$\gamma = -{1/ 3}$ at the critical temperature and
$\gamma = -{1/2}$ off the critical temperature.
We estimate the string
susceptibility at finite area, $\gamma_{\rm est}$, by a suitable
ratio of fixed area
partition sums designed to cancel the cosmological constant:
\beeq
\gamma_{\rm est}=\{\ln ({{\cal Z}_{n+1} {\cal Z}_{n-1}\over {\cal
Z}_{n}^{2}})/\ln (1-{1\over n^{2}})\} +3 \;\; .
\label{gdef}
\eneq
This is a ``local" estimate of $\gamma$ in that it involves
neighbouring partition sums.
Clearly, however, there
are many different ways in which to extract such an estimate.

\begin{figure}

\centerline{
\psfig{file=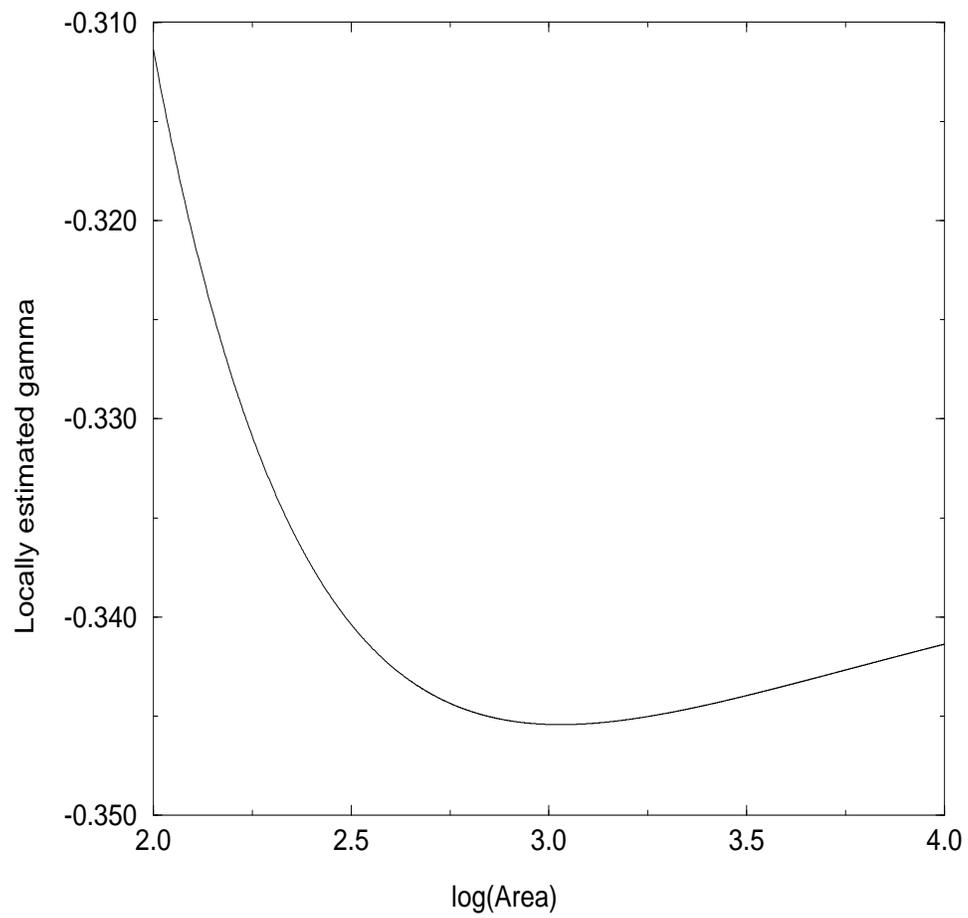,height=14cm,width=14cm,angle=-90}}
\caption
{$\gamma_{\rm est}$ at the critical temperature for quartic
interactions.
}
\label{fig1}
\vspace{0cm}
\end{figure}

We present the results of $\gamma_{\rm est}$ for quartic interactions
at the critical temperature in Fig(\ref{fig1}) and simply note that
the cubic case
demonstrates the same behaviour. Indeed we observe the following
important common features:\newline
(i) For large areas $\gamma_{\rm est}$ approaches the theoretical
expectation both on and off the critical temperature.\newline
(ii) Off the critical temperature $\gamma_{\rm est}$ approaches
$-{1/2}$ rapidly from above. At the critical temperature
$\gamma_{\rm est}$ drops below$-{1/ 3}$ and then slowly converges
towards the theoretical value from below.\newline
(iii) For the region at the critical temperature where
$\gamma_{\rm est}$ is less than $-{1/3}$ the effects of finite
size are greatest in the range approximately bounded by $200 < n <
2000$.

That a qualitative difference on and off the critical temperature
should appear is consistent with the expectation that finite size
effects will be influenced by large Ising spin correlations at the
critical point. However, such a difference has not previously been
taken into account. We observe that
this is, in fact, an important
factor which
cannot be ignored. Significantly, the effects of such finite
size corrections appear most pronounced in regions where previous
numerical
simulations have concentrated their estimates of $\gamma$, which most
typically deal with simulations of the size ranging around $n=1000
\sim 2000${\cite A}. We see that finite
size effects at the critical temperature do not simply decrease with
increasing size but exhibit an important nonlinearity with greatest
effect around the simulation sizes chosen.
In addition we note that, since a range of values for $\gamma_{\rm
est}$ are recovered, it follows that
we can engineer to find many values of $\gamma$.
Indeed, at the critical temperature $\gamma_{\rm est}$ actually passes
through the correct value even for small systems.
It appears then that extraction of $\gamma$
from small area studies can be misleading.

As we pointed out above, these results are exhibited for a particular
choice of form for the estimation of $\gamma$. However, as will be
demonstrated below, while some of the features may change,
our observations in general remain sound.

The magnetic susceptibility results reproduce those expected from
Liouville theory{\cite J} as well as from standard scaling analysis
{\cite F}. In
particular we find that
\beeqar
\chi_{n} &\rightarrow& {\rm constant} \;\;\;\;\;\; ({\rm High \;\;
temp.
\;\; phase}) \nonumber \\
&\rightarrow & n^{2/3}  \;\;\;\;\;\;\;\;\;\; ({\rm on \;\; the \;\;
critical \;\; temp.})\;\; ,
\eneqar
while in the low temperature phase
\beeq
\chi_{n}+n<\sigma >^{2} \rightarrow n \;\; .
\eneq

\section{Singularity Analysis of Finite Size Effects}

We wish now to analytically find the
 asymptotic form of the fixed area partition sums.
This is made possible by the observation
that the large $n$ behaviour of ${\cal Z}_{n}$ is dominated by the
singular points of the free energy.
As before we will deal explicitly with the case of quartic
interactions. We motivate the analysis by applying this approach to
the simple case of pure gravity.

The parametric solution for the one matrix model with quartic
interactions is given by{\cite B}
\beeq
{\cal F}=-{1\over 2}\ln z + {1\over 24}(z-1)(9-z) \;\;\;\; {\rm with}
\;\;\;\;
g(z)\equiv{1-z\over 12z^{2}}= g \;\; .
\label{quadcon}
\eneq
The singularity of $\cal F$ with respect to $g$ is determined by the
condition $g^{\prime}(z_{0})=0$ for which
$z_{0}=2$, so that $g_{c}=-{1/48}$.
The constraint in (\ref{quadcon}) can easily be inverted to yield
\beeq
(z-z_{0})= a(g-g_{c})^{1/2}+b(g-g_{c})+..... \;\; .
\label{zexpand}
\eneq
where for brevity we have not displayed the values of $a,b,...$.
This in turn can be used to  generate an expansion for $\cal F$ about
the
critical point $g_{c}$.
 Being careful to retain sufficient terms
in the expansion the nonregular terms are
found to be
\beeq
{\cal F}(g)\sim {12283{\sqrt 3}\over 5}(g-g_{c})^{5/2}+
{1769472{\sqrt 3}\over 7}(g-g_{c})^{7/2}+......
\label{fexpand}
\eneq

The required asymptotic form can be obtained by employing
the binomial expansion for $(g-g_{c})^{\alpha}$
and subsequently the asymptotic expansion for the gamma function.
Writing
$(g-g_{c})^{\alpha} =\sum_{n} g^{n}{\cal Z}^{\alpha}_{n}$,
one finds
\beeq
{\cal Z}^{\alpha}_{n}\sim -(-g_{c})^{\alpha -n}(-1)^{n}
{\sin (\pi\alpha )\over
\pi}\Gamma (\alpha +1)n^{-(1+\alpha)}\exp (\alpha(1+\alpha)/2n)+.....
\eneq
The term with $\alpha = 5/2$ is seen to
reproduce the leading order behaviour found in (\ref{orig}).
We can also
find the leading order corrections which involve contributions
from both the $(g-g_{c})^{5/2}$ and $(g-g_{c})^{7/2}$ terms:
\beeq
{\cal Z}_{n}\sim {\cal Z}_{n}^{\rm leading}(1+{35\over
8n}-({1769472\over 7})({5\over 12283})({1\over 48})^{-1}
{\Gamma (9/2)\over
\Gamma(7/2)}{1\over n}+....)
\eneq
which is exactly that expressed in (\ref{orig}). We thus have a
method by which
finite size corrections may be derived without relying on a
closed form for $\cal F$.

\begin{figure}

\centerline{
\psfig{file=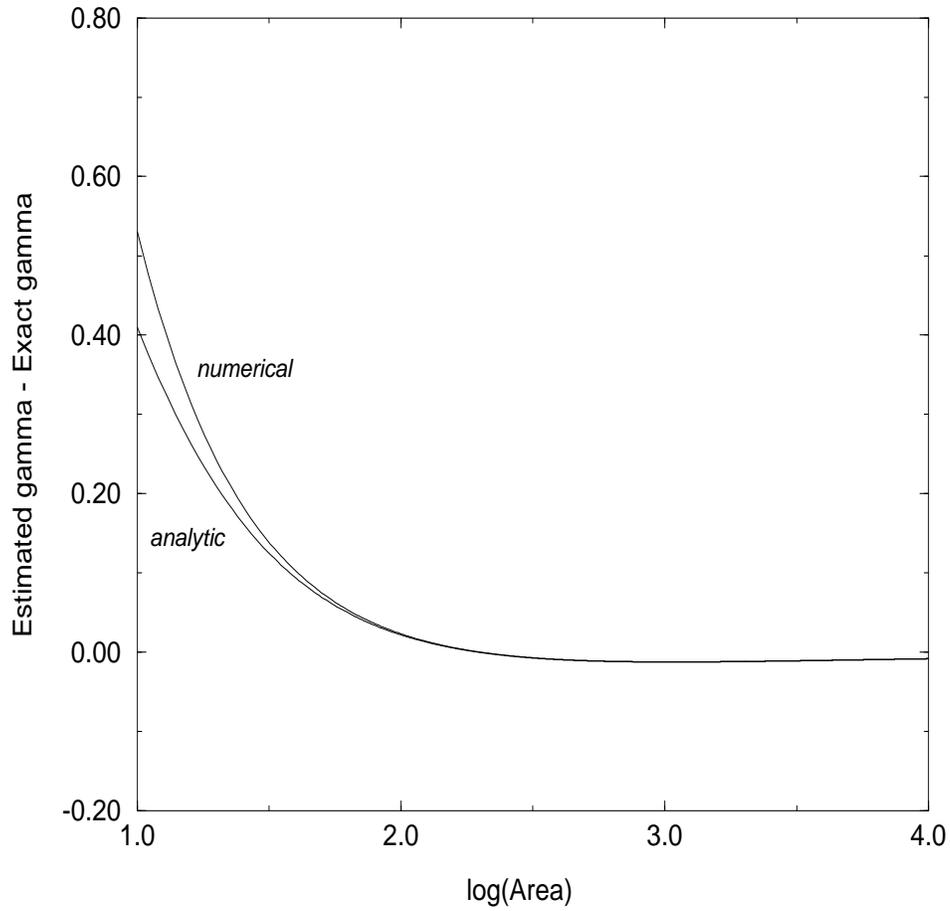,height=14cm,width=14cm,angle=-90}}
\caption
{Comparison of the analytic and numerical estimates of $\gamma$ at the
critical temperature with quartic interactions. The analytic graph
includes all the corrections given in (29).
}
\label{fig2}
\vspace{0cm}
\end{figure}

We now apply this approach to the two matrix model.
{}From the parametric representation of the free energy it follows that
${\cal F}^{\prime}(z_{0})={d{\cal F}/ dz}|_{z_{0}} =0$,
where $z_0$ is given in (\ref{zo}).
The behaviour of the second derivatives with respect to $z$
are:
\beeqar
{\rm Off \; crit. \; temp.}&:&
{\cal F}^{\prime\prime}(z_{0})\not= 0 ,~ g^{\prime\prime}(z_{0})\not=0
\; \; ,\nonumber \\
{\rm On\; crit.\; temp.}&:&
{\cal F}^{\prime\prime}(z_{0})= 0 ,~ g^{\prime\prime}(z_{0})=0\;\; ,
\eneqar
where the non-vanishing of ${\cal F}^{\prime\prime}(z_{0})$ off
criticality
is crucial to obtaining the correct scaling laws.
Hence we differ from{\cite C} in this respect. We must thus
consider two regions.\newline
{\bf (i) The case off the critical temperature, $c\not= c_{\rm crit}$}:
As in the case of pure gravity we invert the constraint
(\ref{im}) to give an
expansion for $(z-z_{0})$ in $(g-g_{c})$:
\beeq
(z-z_{0})=a(g-g_{c})^{1/2}+b(g-g_{c})+.....
\label{zexpand3}
\eneq
Taylor expanding $\cal F$ around $z_{0}$ and
substituting (\ref{zexpand3}) generates a series of both
regular and nonregular terms in $(g-g_c)$. Formally, the lowest
exponent of the
nonregular terms is 3/2 but this term vanishes owing to the relation
\beeq
{\cal F}^{\prime\prime}(z_{0})ab+{a^{3}\over 6}
{\cal F}^{\prime\prime\prime}(z_{0})=0 \;\; ,
\eneq
explicitly requiring that ${\cal F}^{\prime\prime}(z_{0})\not= 0$.
The
nonregular terms contributing to $\cal F$ are thus
\beeq
{\cal F}(g)\sim {\cal A}1(g-g_{c})^{5/2}+{\cal A}2(g-g_{c})^{7/2}+....
\eneq
so that
\beeq
{\cal Z}_{n}\sim (-g_{c})^{-n}n^{-7/2}\{ 1+{{\cal B}1\over
n}+{{\cal B}2\over n^{2}}+...\} \;\; ,
\label{pureexpand}
\eneq
again exhibiting the same basic form as that for pure gravity. Some
representative values for ${\cal B}1$ with quartic interaction are
\beeq
{\cal B}1 = -72.69 \;\; {\rm at} \;\; c=0.20 \;\; ; \;\;
{\cal B}1=-8.76 \;\; {\rm at} \;\; c=0.36 .
\label{paramb}
\eneq
{\bf (ii) At the critical temperature $c=c_{\rm crit}$}:
Since now $g^{\prime\prime}(z_{0})=0$ the expansion of (\ref{im})
takes the form
$(g-g_{c})=(z-z_{0})^{3}{g^{\prime\prime\prime}(z_{0})/
6}+(z-z_{0})^{4}{g^{IV}(z_{0})/ 24}+.......$,
which after inversion gives :
\beeq
(z-z_{0})=a(g-g_{c})^{1/3}+b(g-g_{c})^{2/3}+d(g-g_{c})+.....
\label{zexpand4}
\eneq
Furthermore, since now ${\cal F}^{\prime\prime}(z_{0})=0$, the
Taylor expansion
for $\cal F$ around $z_{0}$ starts at $(z-z_{0})^{3}$.
Substituting (\ref{zexpand4}) into this expansion for $\cal F$
the coefficients
of powers of $(g-g_{c})$ conspire so that the
$(g-g_c)^{4/3}$ and $(g-g_c)^{5/3}$
terms are absent.
 The leading singular behaviour for $\cal F$ is thus given by
${\cal F}\sim (g-g_{c})^{7/3}$.
Consequently, following the same steps as before, the leading
behaviour of ${\cal
Z}_{n}$ will take the form $
{\cal Z}_{n}\sim (-g_{c})^{-n}n^{-10/3}$,
from which it follows that the string susceptibility  is given by
$\gamma=-1/3$. The corrections can be similarly calculated for
which we quote the
results:
\beeqar
{\rm Quartic\;\; interaction} \; : \; {\cal Z}_{n}&=&
{\cal Z}_{n}^{\rm leading}\{
1+{0.4287\over n^{1/3}}-{3.08\over n}-{1.2980\over n^{4/3}}+...\}
\nonumber \\
{\rm Cubic\;\; interaction} \; : \; {\cal Z}_{n}&=&
{\cal Z}_{n}^{\rm leading}\{
1+{0.286\over n^{1/3}}-{3.05\over n}-{0.936\over n^{4/3}}+...\} \;\; .
\label{param}
\eneqar
The crucial observation here is that the next to leading order
correction goes as
$1/ n^{1/3}$ rather than $1/n$ which is the case with pure gravity.
We have verified that these corrections reproduce the
observed finite size effect from the
numerical analysis (see Fig(\ref{fig2})). We have
thus isolated the fundamental difference between these two cases.
It is now clear that
assuming $1/ n$ type corrections both off {\underline {and}} on
the critical temperature
is not justified.

For the magnetic susceptibility
the important expression was given in (\ref{mag})
which we can express as the sum of two integrals
$K(z)=I_{1}+I_{2}$.
We see from (\ref{im2}) that at the critical point $z_{0}$ the
expression
$g(z,H)$ is singular. Consequently both $I_{1}$ and $I_{2}$ are
singular.  However, the sum $I_{1}+I_{2}$ is regular at
$z_{0}$.

As with the free energy we can extract the critical behaviour of the
magnetic susceptibility by taking account of its behaviour about the
singular point.
Expanding $K(z)$ about $z_{0}$ we find that, in the high temperature
phase as
well as at the critical temperature,
$K(z)\sim \alpha (z-z_{0})^{2}+\beta (z-z_{0})^{3}+....$,
while in the low-temperature phase,
$K(z)\sim(z-z_{0})$.
We know from (\ref{zexpand3}) that off the critical temperature
$(z-z_{0})\sim a(g-g_{c})^{1/2}+b(g-g_{c})+...$. Consequently,
the leading
nonregular term is
$K(g)\sim (g-g_{c})^{3/2}$,
where we have verified
that this term does not vanish as was the case with the
free energy for which we know that ${\cal F}\sim (g-g_{c})^{5/2}$.
Expanding in $g^{n}$ it follows that
\beeq
\chi_{n}={K_{n}\over n{\cal Z}_{n}} \rightarrow {\rm constant} \;\; .
\eneq
Similarly at the critical temperature we have $(z-z_{0})\sim
a(g-g_{c})^{1/3}+...$ so that
$K(g)\sim (g-g_{c})^{2/3}$.
We know that ${\cal F}\sim (g-g_{c})^{7/3}$ so that we find
\beeq
\chi_{n}\sim n^{2/3} \;\; .
\eneq
In the low temperature phase, it follows that
$K(g)\sim (g-g_{c})^{1/3}$.
Consequently,
\beeq
{K_{n}\over n{\cal Z}_{n}} \sim n \;\; .
\eneq

We can thus account for the behaviour expected
from standard Fisher-scaling theory{\cite F}
as well as from Liouville theory{\cite J}.

\section{Alternate Estimates of $\gamma$}

We now investigate
alternate definitions for the
estimation of $\gamma$. From ${\cal Z}_{n}$
one can introduce an obvious such
alternative:
\beeq
\gamma_{\rm alt.\; est}=\gamma_{\rm exact}+{\ln (1+{\rm finite\; size\;
corrections})\over \ln (n)} \;\; ,
\eneq
from which it follows that
\beeqar
c&=&c_{\rm crit}:
 \gamma_{\rm alt.\; est}\sim \gamma_{\rm exact}
+{c_{1}\over \ln (n)n^{1/3}}
\;\; ; \;\; \nonumber \\
c&\not=& c_{\rm crit}:\gamma_{\rm alt.\; est}
\sim \gamma_{\rm exact}+{a_{1}\over \ln (n)n} \;\;,
\label{gapprox}
\eneqar
where $c_{1}>0$ and $a_{1}<0$.
According to this definition  the finite size behaviour predicted is
opposite that found from the numerical results for
$\gamma_{\rm est}$. That is, off the critical temperature
$\gamma_{\rm alt. \; est}$ approaches $-1/2$ from below while on the
critical temperature it approaches $-1/3$ from above.
There is no real inconsistency here as  there are  many ways
in which to estimate $\gamma$, and the different estimates are
only required to coincide asymptotically.
For example,
Brezin and Hikami{\cite I} use an estimate based on
Pad\'e approximation where
\beeq
\gamma_{\rm Pade \; est}=3-{n(1+n)(f_{n}-f_{n-1})\over
(1+n)f_{n}-nf_{n-1}}
\;\; ,
\label{pade}
\eneq
and $f_{n}$ is a Pad\'e approximant to the ratio ${\cal
Z}_{n}/{\cal Z}_{n-1}$. We observe that each choice has a
{\underline{different}} finite size behaviour. We can explicitly
demonstrate this where in addition to those results in
(\ref{gapprox}) we have that on the critical temperature, where ${\cal
Z}_{n}\sim {\cal Z}_{n}^{\rm leading}(1+c_{1}/n^{1/3}+c_{2}/n+...)$
and $c_{1}>0$,
\beeqar
\gamma_{\rm
est}&\sim&-{1\over 3}-{4c_{1}\over 9n^{1/3}}+{5c_{1}^{2}\over
9n^{2/3}}-{2c_{1}^{3}/3+c_{2}\over n}+.....\;\;\;\;{\rm and}
\nonumber \\
\gamma_{\rm Pade\; est}&\sim&-{1\over 3}-{4c_{1}\over 9n^{1/3}}+
{5c_{1}^{2}\over 9n^{2/3}}+{-2c_{1}^{2}/3-2c_{2}-3b+b^{2}\over n}+....
\eneqar
where $b=3-\gamma $,
while off the critical temperature where ${\cal Z}_{n}\sim {\cal
Z}_{n}^{\rm leading}(1+a_{1}/n+a_{2}/n^{2}+...)$ and $a_{1}<0$,
\beeqar
\gamma_{\rm
est}&\sim&-{1\over 2}-{2a_{1}/n}+{3a_{1}^{2}-6a_{2}\over n^{2}}
+.... \;\;\;\; {\rm and} \nonumber \\
\gamma_{\rm Pade\; est}&\sim&-{1\over 2}-{2a_{1}+3b-b^{2}\over n}+....
\eneqar
where we note that $-2a_{1}-3b+b^{2}$ is
positive and that $\gamma_{\rm est}$ is in fact analytically consistent
with the numerical results.
The situation is thus similar to the scheme dependence in
renormalisation. {\bf Any particular prescription for $\gamma$ is
suitable
but comparison of different definitions is not a meaningful exercise.}
Thus different models should be compared using the same definition for
$\gamma_{\rm est}$.
{}From these analyses we can summarise the general structure of finite
size corrections pertinent to any choice of estimate:\newline
(i) They are not always parameterisable as $1+{\alpha}/n+...$
.\newline
(ii) The parameterisation is dependent on the
expected value of $\gamma$.\newline
(iii) If $\gamma_{\rm expected}=q/p$ for
$q,p\in {\bf Z}$ then in general the finite size corrections will be
parameterised as $1+{\alpha}/n^{1/p}+....$
This is our  general ansatz for the form of finite size
corrections.
\newline
(iv) Some of the coefficients in the expansion may be zero, for
instance in the Ising case at the critical temperature we have the
expansion $1+c_{1}/n^{1/3}+c_{2}/n+c_{3}/n^{4/3}+c_{4}/n^{2}+...$

\section{Some Comments on Minbu Analysis}

A particular
approach to extracting the string susceptibilty from
numerical simulations
is given by measuring the distribution of minimum
neck baby universes (minbus) on random surfaces of a given fixed total
area $A$. An estimation for the average number of minbus of
size $B$, ${\overline n}_{A}(B)$, was given by Jain and Mathur
{\cite E} and
takes the form
\beeq
{\overline n}_{A}(B)={3(B+1)(A-B+1){\cal Z}_{B+1}{\cal Z}_{A-B+1}\over
{\cal Z}_{A}} \;\; .
\label{mindis}
\eneq
For quartic interactions
we should substitute 4 for 3. It follows
that (to leading order)
\beeq
\ln ({\overline n}_{A}(B))={\rm constant}+(\gamma -2)\ln (B(1-{B\over
A})) \;\; .
\label{minmes}
\eneq
Thus by measuring the slope of this function we can numerically
determine $\gamma${\cite A}.

\begin{figure}

\centerline{
\psfig{file=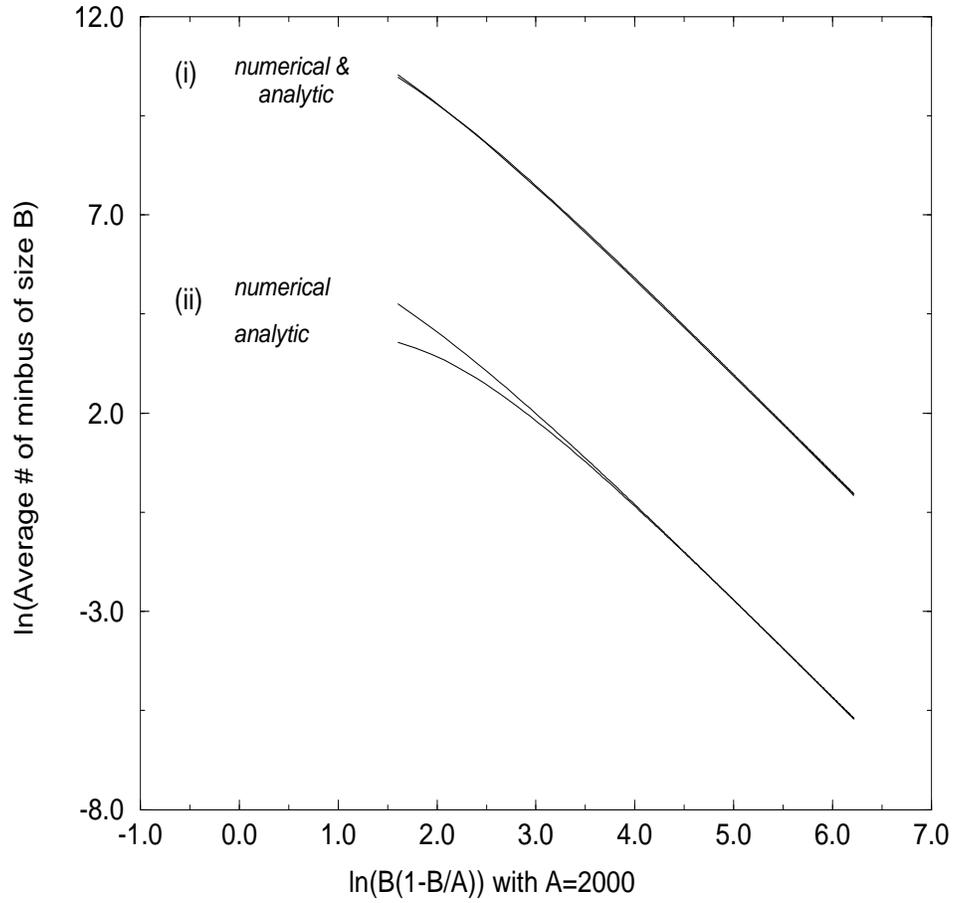,height=14cm,width=14cm,angle=-90}}
\caption
{
Numerically and analytically derived minbu plots for (i) pure gravity
and (ii) the Ising case off the critical temp. ($c=0.36$). The
calculated values of $\ln ({\overline n}_{A}(B))$  in (ii) have been
shifted by a
constant for clarity. All are with quartic interactions.}
\label{fig3}
\vspace{0cm}
\end{figure}

The form of finite size corrections
for pure gravity being
$(1+\alpha/n+...)$
it follows that the leading
finite size correction to (\ref{minmes}) is
\beeq
\ln ({\overline n}_{A}(B))={\rm constant}+(\gamma -2)\ln (B(1-{B\over
A}))+(\gamma-2+\alpha)\ln (1+{1\over B(1-B/A)})+ ... \;\; ,
\label{minmes2}
\eneq
We know from (\ref{orig})
that for pure gravity with
quartic interactions the finite size parameter has the value
$\alpha=-25/8$. Plotting the minbu distribution (\ref{minmes2}) and the
exact minbu distribution calculated using (1) in Fig(\ref{fig3}i)
we see that
this correction gives excellent agreement with the exact result.
Applying this approach to the Ising case off the critical temperature,
where the finite size corrections were given in (\ref{paramb})
for the high and low
temperature phases, we find that these finite size corrections
{\underline{do not}} accurately mimic the exact minbu distributions,
Fig(\ref{fig3}ii).
Clearly, higher order corrections are  more
important in the Ising case.
This behaviour is again evident at the critical temperature, where we
must also account for $1/n^{1/3}$ type corrections.
Again we find that retaining higher order corrections is necessary in
order to
obtain a reasonable fit.

Since minbus are measured over a range of volumes finite
size effects are unavoidable for small minbus.
As these finite size effects are hard to extract for cases
where either $\gamma$ is not known or where $\gamma = q/p $ with
large integer $p$,
it appears that
applying the minbu technique to
extract meaningful estimates
for $\gamma$ from simulations in
these interesting cases is fraught with
difficulties.

\section{Conlusion}

A simple parameterisation of finite size effects is a natural first
approach to analysing  numerical
simulations. As we have shown, however,
the actual parameterisation is nontrivially
dependent on the exact value of the string susceptibility. On
this basis we have proposed a general ansatz for the form of
finite size
corrections.
A
possible algorithm then is to make a best guess for $\gamma$ and to
fit this
to the observed data with the finite size corrections corrrectly
included. By performing a $\chi^{2}$ analysis the best fit for a
particular $\gamma$ could be recursively searched.

We have compared different approaches to estimating $\gamma$ and
demonstrated that attempts to compare estimates from different
definitions can be misleading.
These considerations become relevant if we wish to extract
reliable numerical
estimates beyond the $c_{M}=1$ barrier
where it
is known that large logarithmic corrections also play a role. A
clear understanding of the functional form for the finite area
estimates
are indispensable in these cases.
\section{Acknowledgements}
It is a pleasure to thank N. Ishibashi for enlightening discussions
and for carefully reading the manuscript. We are also grateful to
H. Kawai and N. Tsuda for many
helpful discussions. H. D. acknowledges the Ministry of Science,
Culture and Education for financial support. The support of JSPS and
the Australian Academy of Science is also gratefully acknowledged by B.
E. H.

\end{document}